\documentclass{emulateapj}

\newcommand{\pfrac}[2]{\left(\frac{#1}{#2}\right)}
\def\eps{\epsilon}

\shorttitle{Residual collisions in GRB outflows} \shortauthors{Li \& Waxman}

\begin{document}

\title{Prompt optical emission from residual collisions in GRB outflows}

\author{Zhuo Li\altaffilmark{1} and Eli Waxman\altaffilmark{1}}

\altaffiltext{1}{Physics Faculty, Weizmann Institute of Science, Rehovot 76100, Israel}

\begin{abstract}
The prompt $\gamma$-ray emission in $\gamma$-ray bursts is believed to be produced by internal
shocks within a relativistic unsteady outflow. The recent detection of prompt optical
emission accompanying the prompt $\gamma$-ray emission appears to be inconsistent with this
model since the out flowing plasma is expected to be highly optically thick to optical
photons. We show here that fluctuations in flow properties on short, $\sim1$~ms, time scale,
which drive the $\gamma$-ray producing collisions at small radii, are expected to lead
to "residual" collisions at much larger radii, where the optical depth to optical photons is low.
The late residual collisions naturally account for the relatively bright optical emission.
The apparent simultaneity of $\gamma$-ray and optical emission is due to the highly relativistic
speed with which the plasma expands. Residual collisions may also account for the X-ray emission
during the early "steep decline" phase, where the radius is inferred to be larger than the
$\gamma$-ray emission radius. Finally, we point out that inverse-Compton emission from residual collisions at
large radii is expected to contribute significantly to the emission at high energy, and may therefore
"smear" the pair production spectral cut-off.
\end{abstract}


%%%%%%%%%%%%%%%%%%%%%%%%%%%%%%%%%%%%%%%%%%%%%%%%%%%%%%%%%%%%%%%%
\section{Introduction}
In the leading model for $\gamma$-ray bursts (GRBs), the energy source is a compact
object that drives a relativistic unsteady outflow with fluctuating Lorentz factor.
Internal shocks within the outflow dissipate the bulk kinetic energy and produce
$\gamma$-rays \citep{RM94,PX94}. A substantial fraction of the outflow kinetic energy may
be dissipated in this model outside the photosphere, allowing one to account for the
non-thermal spectra and for the complicated light curves of GRBs \citep{Kobayashi97}. The
internal shocks are expected to generate/amplify magnetic fields and to accelerate
electrons, which produce MeV $\gamma$-rays by synchrotron emission \citep[For review of
internal shock models, see][]{Waxman rev}.

Due to the relatively short duration of the prompt $\gamma$-ray emission,
$T\sim1$~to~$10^2$~s, the observation of long-wavelength (optical) prompt emission is a
difficult task. GRB 990123 was the first event for which optical emission was detected
during the burst \citep{Akerlof99}. Today, thanks to the rapid localization of GRBs by
the {\it Swift} satellite \citep[see][for recent review]{Zhang07}, a larger number of
optical (and longer wavelengths) observations are carried out during the bursting phase
\citep[e.g.][]{Blake05,Vestrand05,Vestrand06,Yost07}.

As shown in \S~\ref{sec:absorption} \citep[see also][]{LZ}, during the emission of prompt
$\gamma$-rays the plasma is expected to be optically thick to optical photons due to
strong synchrotron self-absorption. This appears to be inconsistent with the detection of
bright optical emission accompanying $\gamma$-ray emission. We point out here that the
internal collisions at small radii, which produce the $\gamma$-ray emission, are expected
to lead to "residual" collisions at much larger radii, where the optical depth to optical
photons is low. These late residual collisions may naturally account for the relatively
bright optical emission. The apparent simultaneity of $\gamma$-ray and optical emission
is due to the highly relativistic speed with which the plasma expands. The time delay
between $\gamma$-ray and optical emission is expected to be shorter than a second, too
short to be identified by current optical observations which usually have lower temporal
resolution. We discuss in \S \ref{sec:large radii} residual collisions in unsteady
outflows, and derive the long-wavelength emission they are expected to produce at large
radii. The implications of our results are discussed in \S~\ref{sec:discussion}.

\section{Strong synchrotron absorption at small radii}
\label{sec:absorption}

We first show that during the prompt $\gamma$-ray emission phase the plasma is highly
optically thick to optical photons. Consider a relativistic outflow with a Lorenz factor
fluctuating over a timescale $t_{\rm var}$. We denote the mean Lorentz factor by
$\Gamma$, its variance by $\sigma^2_{\Gamma}$, and assume $\sigma_{\Gamma}\la\Gamma$. The
internal collisions that produce $\gamma$-rays occur in this model at a radius
$R_\gamma\sim\Gamma^3ct_{\rm
var}/\sigma_\Gamma\approx10^{12.5}(\Gamma/\sigma_\Gamma)\Gamma_{2.5}^2t_{{\rm
var},-3}$cm, where $\Gamma_{2.5}=\Gamma/10^{2.5}$, and $t_{{\rm var},-3}=t_{\rm
var}/10^{-3}$s. The observed variability time implies
$R_\gamma\lesssim10^{14}\Gamma_{2.5}^2$~cm for a large fraction of BATSE bursts
\citep{Woods95}. We assume that internal collisions lead to shocks that generate/amplify
magnetic fields and accelerate electrons to high energy, leading to synchrotron emission
that accounts for the prompt $\gamma$-rays.

For typical outflow parameters, the cooling time of the electrons, $t_c$, is short
compared to the dynamical time, $t_d$, over which the plasma expands. The dynamical time
measured in the plasma frame is $t_d\sim R/\Gamma c\sim1R_{13}/\Gamma_{2.5}$~s, where
$R_{13}=R_\gamma/10^{13}$~cm. The cooling time of electrons with Lorentz factor
$\gamma_\nu$, emitting synchrotron photons of frequency $\nu$, is $t_c\approx\gamma_\nu
m_ec^2/P_{\rm syn}(1+y)$, where $P_{\rm syn}=(4/3)\sigma_Tc\gamma_\nu^2B^2/8\pi$ and
$y\equiv P_{\rm IC}/P_{\rm syn}$ is the ratio between inverse-Compton and synchrotron
emission, which is roughly given by the ratio of radiation and magnetic field energy
densities, $U_\gamma/(B^2/8\pi)$. In order to account for the $\gamma$-ray emission, the
magnetic field energy density needs to be close to equipartition with the thermal energy
of the plasma. Since a significant fraction of this thermal energy is emitted as
$\gamma$-rays, we estimate $B^2/8\pi\sim U_\gamma$. Using $U_\gamma\simeq L_{\gamma}/4\pi
R_\gamma^2\Gamma^2c$ this gives
$B\sim10^5L_{\gamma,51}^{1/2}\Gamma_{2.5}^{-1}R_{13}^{-1}$~G and
$t_c\sim10^{-2}\Gamma_{2.5}/B_5\nu'_{15}(1+y)$s, where $L_{\gamma,51}=L_\gamma/10^{51}\rm
erg\,s^{-1}$, $B_5=B/10^5$~G, and $\nu'_{15}=\nu'/10^{15}$~Hz is the observed frequency,
$\nu'=\Gamma\nu=\Gamma\gamma^2_\nu eB/2\pi m_ec$. Since $t_c(\nu'_{15}=1)\ll t_d$, the
electrons, which were initially accelerated to high energy at which their synchrotron
emission peaks at $\sim1$~MeV, rapidly cool down to energies at which their synchrotron
emission peaks well below the optical band. Neglecting synchrotron self-absorption, this
would have lead to a synchrotron spectrum of $F_\nu\propto \nu^{-1/2}$ extending from the
$\gamma$-ray band to below the optical band ($F_\nu$ stands for the flux per unit
frequency).
%$\gamma_c=0.07\Gamma_{2.5}B_5^{-2}R_{13}^{-1}(1+y)^{-1}$

Self-absorption of photons of frequency $\nu$ is dominated by electrons with Lorenz factor
$\gamma_\nu$, which constitute a fraction $t_c(\gamma_\nu)/t_d$ of the electron population.
We may therefore approximate the (volume averaged) absorption coefficient by $\alpha_\nu\approx
n_e[t_c(\gamma_\nu)/t_d] e^3B /2\gamma_\nu(m_ec\nu)^2$, where the electron density is given by
$n_e=L_k/4\pi \Gamma^2R^2m_pc^3$, with $L_k$ the kinetic luminosity of the GRB outflow.
The self-absorption frequency, where the optical depth $\alpha_\nu R/\Gamma$ equals unity, is
\begin{equation}\label{eq:absorp_nu}
  h\nu'_a\approx0.3L_{k,52}^{1/3}\Gamma_{2.5}^{1/3}R_{13}^{-2/3}(1+y)^{-1/3}\rm keV,
\end{equation}
independent of $B$, and the corresponding electron Lorenz factor is
$
  \gamma_a\equiv\gamma_\nu(\nu_a)\approx30L_{k,52}^{1/6}\Gamma_{2.5}^{-1/3}B_5^{-1/2}R_{13}^{-1/3}(1+y)^{-1/6}.
$ Here $L_{k,52}=L_k/10^{52}\rm erg\,s^{-1}$. Note, that the electron cooling rate is
modified below $\gamma_a$, and $t_c$ becomes larger than that used for deriving
eq.~(\ref{eq:absorp_nu}), due to the absorption of radiation. However this modification
is not large for $y\sim1$, in which case cooling by inverse-Compton emission is
comparable to synchrotron cooling.

Examining eq.~(\ref{eq:absorp_nu}), we expect a large optical depth below the X-ray band
and hence a strong suppression of the optical flux. This appears to be inconsistent with
observations, which typically show $F_{\nu_{\rm op}}\ga F_{\nu_\gamma}$
\citep[e.g.,][]{Yost07}. It should be mentioned here that, within the context of the
current model, the constraint $R_\gamma<10^{14}$~cm, which implies $\nu'_a\gg1$~eV, is
obtained not only from the observed variability time, $t_{\rm var}$, but also from the
requirement that the synchrotron emission peaks in the MeV band. The characteristic
(plasma frame) Lorentz factor of the $\gamma$-ray emitting electrons is $\gamma_e\sim
m_p/m_e$ (see \S~\ref{sec:rad}), leading to synchrotron emission peaking at
\begin{equation}\label{eq:peak_nu}
  h\nu'_p\approx\hbar\Gamma\gamma_e^2\frac{eB}{m_ec}
  \approx0.3L_{\gamma,51}^{1/2}R_{13}^{-1}\rm MeV.
\end{equation}
$h\nu'_p\sim1$~MeV implies therefore $R_\gamma\lesssim10^{13}$~cm. This constraint may be
avoided, for bursts where $R_\gamma\lesssim10^{13}$~cm can not be inferred from $t_{\rm var}$,
in a model where $\gamma$-ray emission is assumed to be produced by inverse-Compton scattering
of $h\nu'_p\ll1$~MeV synchrotron photons (assuming magnetic field well below equipartition).
In such a model the inverse-Compton spectrum is expected to be hard, $F_\nu\propto\nu^2$, at low
frequencies, $h\nu'<1$~MeV, due to self-absorption of the synchrotron spectrum
\citep[e.g.][]{Pana00}. The observed spectrum is softer for most bursts.

\section{Large-radius emission from residual collisions}
\label{sec:large radii}

The optical depth for optical photons drops below unity at radii $R\ga10^{15}$cm (see eq.
\ref{eq:absorp_nu}; note $(1+y)\propto R^{2/3}$, see \S\ref{sec:rad}). We show here that
the optical emission could be produced by "residual" collisions at such large radii.
Note, that the time delay between $\gamma$-ray and optical emission in this model,
\begin{equation}\label{eq:delay time}
  \tau_{\rm delay}\approx R_{\rm op}/2\Gamma^2c\sim0.2R_{\rm op,15}\Gamma_{2.5}^{-2} \rm s,
\end{equation}
is expected to be shorter than the characteristic temporal resolution of the optical
observations, which is a few seconds. Thus, optical and $\gamma$-ray emission may appear
to be simultaneous. However, better temporal resolution may allow one to detect a
systematic time delay between the two wave bands. In addition, one would expect larger
observed variability timescales at longer wavelengths, $t_{\rm var, op}\sim\tau_{\rm
delay}$.

We approximate the outflow by a sequence of $N\gg1$ equal mass shells ($i=1,\ldots,N$)
separated by an initial fixed distance $ct_{\rm var}$ and expanding with (initial) Lorenz
factors $\Gamma_{i,0}$ drawn from a random distribution with an average $\Gamma$ and
initial variance $\sigma^2_{\Gamma,0}<\Gamma^2$. We assume that the radial extent of the
outflow $Nct_{\rm var}$ is much smaller than the collision radii $R>\Gamma^2 ct_{\rm
var}$, i.e. $N\ll\Gamma^2$, which is reasonable given the observed variability
\citep[e.g][]{fm95}. The model may, of course, be complicated, e.g. by adding several
variability times or by allowing variable mass shells. Adding such degrees of freedom may
allow one to control the details of the predicted long wave length emission. Our main
goal is to demonstrate that the simplest model considered here may naturally account for
the observed optical emission.

The dynamics of late residual collisions is discussed in \S~\ref{sec:dynamics}, and the
radiation they are expected to generate is discussed in \S~\ref{sec:rad}.

\subsection{Late residual collisions}
\label{sec:dynamics}

Let us first consider the evolution of the outflow using the simplifying assumption that
shells merge after collisions. This assumption would be approximately valid if all the
internal energy generated by a collision of two shells is radiated away. As the flow
radius increases, the typical number $n(R)$ of initial shells that merge into one single
shell increases, and the variance of the Lorenz factors of the resulting shells
decreases. For a group of shells with a small Lorenz factor variance, the velocities
$v_i$ of the shells in the shells' center of momentum frame are not highly relativistic.
In this case, conservation of momentum implies that the velocity of a merged group of
shells is given by the average of merged shells' velocities, $\bar v=(1/n)\sum_{i=1}^n
v_i$, and that the variance of the velocities of merged groups of shells is
$\sigma_v(n)=\sigma_{v,0}/\sqrt{n}$ where $\sigma_{v,0}$ is the initial variance. This,
in turn, implies that the variance of (observer frame) Lorenz factors,
$\sigma_\Gamma(n)/\Gamma\approx\sigma_{v}(n)/c$, evolves like
$\sigma_\Gamma(n)=\sigma_{\Gamma,0}/\sqrt{n}$. Collisions of merged groups of $n$ shells
will therefore take place at a radius $R(n)\sim\Gamma^3c\times nt_{\rm
var}/\sigma_\Gamma(n)$, which implies
\begin{equation}\label{eq:residual n&sigma}
  n\propto R^{2/3}, ~~\sigma_{v}\propto\sigma_\Gamma\propto R^{-1/3}.
\end{equation}
The outflow energy that may be dissipated and radiated away is the energy associated with
the random velocities of the shells (in the outflow rest frame). This energy decreases as
\begin{equation}\label{eq:residual energy}
  E_{\rm fluc}\propto\Gamma \sigma_{v}^2\propto R^{-2/3}.
\end{equation}

\begin{figure}
\hskip-.5cm\includegraphics[width=10cm]{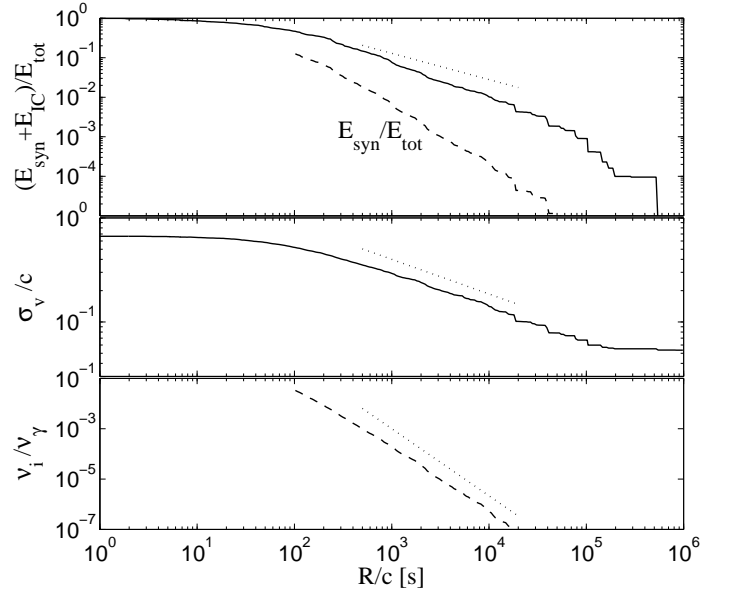} \caption{ The evolution of an outflow
composed of $N=10^3$ equal mass shells initially separated by $ct_{\rm var}=c\times1$~ms,
with random Lorenz factors $\Gamma_i=10^{2.5}\times 3^\xi$ where $\xi$ is normally
distributed with zero mean and unit standard deviation. In each collision it is assumed
that 1/3 of the internal energy generated is radiated (see \S~\ref{sec:dynamics} for more
details).  {\em Top panel}: The energy radiated by collisions at radii larger than $R$,
normalized to the total radiated energy (solid line- synchrotron and IC emission, dashed
line- synchrotron emission only); {\em Middle panel}: The standard deviation of shell
velocities (in the center of momentum frame) as function of $R$; {\em Bottom panel}: The
characteristic frequency of synchrotron radiation (see \S~\ref{sec:rad}) as a function of
$R$, normalized to its value at the smallest collision radius. Dotted lines show the
approximate analytic scaling laws obtained for the "merging-shell" model
(eqs.~\ref{eq:residual n&sigma},~\ref{eq:residual energy},~ \ref{eq:residual nu_i}).
$\sim1$~\% of the radiated energy is released at a radius $R/c\sim10^4$~s, where residual
collisions produce optical synchrotron photons, $\nu_i/\nu_\gamma\sim10^{-6}$. Most of
this energy is emitted by inverse-Compton scattering of the prompt $\gamma$-rays.}
\label{fig}
\end{figure}

Let us consider next the evolution of the outflow dropping the assumption of shell
merger. In order to describe the evolution in this case we carried out a numerical
simulation, assuming that in each collision one third of the kinetic energy of the two
shells (in the center of momentum frame) is radiated away, and that the shells separate
after the collision, each carrying half the remaining energy (in the center of momentum
frame). Fig. \ref{fig} shows the evolution of an outflow with the following parameters:
$N=10^3$, $t_{\rm var}=1$~ms and $\Gamma_i=10^{2.5}\times 3^\xi$ with $\xi$ normally
distributed with zero mean and unit standard deviation. As can be seen from the top two
panels of the figure, the evolution of $\sigma_v$ and of the radiated energy are well
approximated by the analytic expressions of eq.~(\ref{eq:residual n&sigma})
and~(\ref{eq:residual energy}), which were obtained under the shell merger assumption. We
will therefore use these approximate analytic expressions in the next section, where the
emitted radiation is discussed.

\subsection{Predicted emission}
\label{sec:rad}

Let us first consider the energy band into which energy is radiated.
We make the common assumptions, that internal shocks
accelerate electrons and generate or amplify magnetic fields,
such that the post-shock electrons and magnetic fields carry fixed
fractions, $\eps_e$ and $\eps_B$ respectively, of the post-shock internal energy.
Under these assumptions, the characteristic
Lorentz factor of post-shock electrons (in the outflow co-moving frame)
scales as $\gamma_i\propto\eps_e\sigma_{v}^2$, and the post-shock magnetic field scales as
$B^2\propto\eps_B\sigma_{v}^2n_e$ (the particle number density scales as
$n_e\propto R^{-2}$). Using eq.~(\ref{eq:residual n&sigma}), the characteristic (observer frame)
frequency of synchrotron photons, $\nu_i\propto\Gamma\gamma_i^2B$, scales as
\begin{equation}\label{eq:residual nu_i}
  \nu_i\propto\sigma_{v}^5R^{-1}\propto R^{-8/3}.
\end{equation}
As can be seen in the bottom panel of fig.~\ref{fig}, eq.~(\ref{eq:residual nu_i}), which
is based on the analytic approximations of eq.~(\ref{eq:residual n&sigma}) for the
simplified "merging-shell" model, describes well also the results of the numerical
simulation for the non-merging model.

Next, consider the emitted flux. It is straight forward to show that the cooling time of
the electrons is short compared to the dynamical time during the late residual collision
phase, up to radii $R\sim10^3 R_\gamma$. We therefore assume that electrons radiate away
all their energy. During the phase of late residual collisions, the plasma is immersed in
the radiation bath of the prompt $\gamma$-rays. The radiation energy density dominates
the magnetic field energy density, since the photon energy density drops as
$U_\gamma\propto R^{-2}$ and the ratio
$y=U_\gamma/(B^2/8\pi)\propto\sigma_{v}^{-2}\propto R^{2/3}$ increases with $R$. The
electrons lose therefore most of their energy by IC cooling, and only a fraction
$1/(1+y)\approx y^{-1}\propto R^{-2/3}$ of the radiated energy is emitted as synchrotron
radiation. Neglecting synchrotron self-absorption, the observed (time-integrated)
spectrum would be $\nu F_\nu\propto E_{\rm fluc} y^{-1}|_{\nu_i=\nu}\propto
R^{-4/3}|_{\nu_i=\nu}\propto\nu^{1/2}$.

Finally, let us consider the effects of synchrotron self-absorption. Eq.
(\ref{eq:absorp_nu}) is valid for $\nu_a<\nu_i$, which implies $\nu_a\propto
R^{-2/3}y^{-1/3}\propto R^{-8/9}$. For $\nu_a<\nu_i$ we therefore have
$\nu_a/\nu_i\propto R^{16/9}$, implying that the optical depth to synchrotron photons
emitted by electrons with the characteristic Lorentz factor $\gamma_i$ will exceed unity
at sufficiently large radii, $R>R_{ia}$. Since at small radii, $R\sim R_\gamma$,
$\nu_i=\nu_\gamma\sim1$~MeV and $\nu_a\sim0.3$~keV (see eq.~\ref{eq:absorp_nu}),
$\nu_i=\nu_a$ is obtained at $R=R_{ia}\sim 10^2R_\gamma$. At this radius
$\nu_i=\nu_a=\nu_{ia}\equiv\nu_i(R_{ia})\sim10$~eV. Thus, the $\nu F_\nu\propto\nu^{1/2}$
(time-integrated) spectrum obtained above neglecting self-absorption does not extend down
to the optical band. In order to derive the spectrum at lower frequencies,
$\nu<\nu_{ia}$, we first derive the evolution of $\nu_a$ at $R>R_{ia}$. For these radii
one needs to consider the electrons accelerated to Lorentz factors larger than the
characteristic Lorentz factor $\gamma_i$, since these electrons dominate emission and
absorption at $\nu>\nu_i$. Shock acceleration is expected to generate a power-law energy
distribution of electrons, $dn_e/d\gamma_e\propto\gamma_e^{-2}$ at $\gamma_e>\gamma_i$.
For this energy distribution, the volume averaged number density of electrons with
Lorentz factor $\gamma_\nu>\gamma_i$ is
$n_e[t_c(\gamma_\nu)/t_d](\gamma_i/\gamma_\nu)=n_e[t_c(\gamma_\nu)/t_d](\nu_i/\nu)^{1/2}$.
Using the same argument leading to eq.~(\ref{eq:absorp_nu}) we find, for $\nu_a>\nu_i$,
\begin{equation}\label{eq:residual nu_a}
  \nu_a\propto \sigma_{v}^{9/7}R^{-5/7}\propto R^{-8/7}.
\end{equation}
The flat electron energy distribution, $\gamma_e^2dn_e/d\gamma_e\propto\gamma_e^0$,
generates equal amounts of synchrotron energy in logarithmic photon energy
intervals, $\nu F_\nu\propto\nu^0$ for $\nu>\nu_a$ (when $\nu_a>\nu_i$).
We therefore obtain for $\nu<\nu_{ia}$
a (time integrated) spectrum given by $\nu F_\nu\propto E_{\rm
fluc}y^{-1}|_{\nu_a=\nu}\propto R^{-4/3}|_{\nu_a=\nu}\propto\nu^{7/6}$.

Combining the above results, the observed flux at $\nu<\nu_{ia}$ is given by
\begin{equation}\label{eq:F_ratio}
\frac{F_\nu}{F_{\nu_\gamma}}\simeq\pfrac{\nu_{ia}}{\nu_\gamma}^{-1/2}\pfrac{\nu}{\nu_{ia}}^{1/6}
\sim10^{2}\pfrac{h\nu'}{\rm
  1\,eV}^{1/6}.
\end{equation}

Several comments should be made here. The flux ratio given by eq.~(\ref{eq:F_ratio})
holds only on average. The observed flux ratios in individual GRB events may differ
significantly, since for a small number of shells (and collisions) large variations in
the late residual collisions should be expected. It should also be noticed that we have
assumed $\sigma_{\Gamma,0}<\Gamma$, while initial conditions with
$\sigma_{\Gamma,0}>\Gamma$ may lead to more efficient $\gamma$-ray production at small
radii, in which case $F_\nu/F_{\nu_\gamma}$ should be smaller by a factor of a few than
the ratio given in eq.~(\ref{eq:F_ratio}).

\section{Discussion}
\label{sec:discussion}

We have shown that late residual collisions, that occur at radii much larger than those
where $\gamma$-ray producing collisions take place, may naturally account for the
observed strong optical emission accompanying the prompt GRB. Internal collisions at
small radii reduce the variance of colliding shell velocities. As a result, the energy
available for radiation at large radii and the characteristic frequency of radiated
photons decrease with radius. We find that one may expect optical to $\gamma$-ray energy
ratio $\sim10^{-4}$, with large burst-to-burst scatter (see fig.~\ref{fig},
eq.~\ref{eq:F_ratio} and discussion at the end of \S~\ref{sec:rad}). This is consistent
with the results of \cite{Yost07}, who find that during the prompt emission of GRBs the
spectral indices between the optical and the $\gamma$-ray bands are in the range of
$0<\beta_{{\rm op}-\gamma}<0.5$, corresponding to $F_{\nu_{\rm
op}}/F_{\nu_\gamma}\sim1-10^3$ and implying that the optical emission is only a small
fraction, $\sim10^{-6}-10^{-3}$, of the total emitted energy.

Although the optical emission is produced at large radii, where synchrotron
self-absorption is avoided, the expected time delay between $\gamma$-ray and optical
emission, $\sim0.1$~s (see eq.~\ref{eq:delay time}), is shorter than the characteristic
temporal resolution of the optical observations, which is a few seconds
\citep[e.g.][]{TZ06}. Thus, optical and $\gamma$-ray emission may appear to be
simultaneous. However, better temporal resolution may allow one to detect a systematic
time delay between the two wave bands. In addition, one would expect larger observed
variability timescales at longer wavelengths, $t_{\rm var, op}\sim\tau_{\rm delay}$.

\citet{Wei07} has suggested that optical emission may be generated by strong internal
shocks at radii $R/c>10^6$~s, driven by shells emitted with a large time delay,
$\sim10$~s, following those producing the main $\gamma$-ray emission \citep[see
also][]{Fan05}. Our model is quite different. We show that optical emission is naturally
expected to arise, without postulating the existence of delayed shells, by residual
collisions at $R/c\sim10^4$~s, in which the characteristic emitted photon frequency is
low, $h\nu\sim1$~eV, due to the reduction of the Lorentz factor variance in the flow
(rather than by the large radius $R/c>10^6$~s). Moreover, we have shown that the optical
luminosity can be estimated from the burst $\gamma$-ray properties, and that it is
consistent with the observations.

The energy released in residual collisions of the relativistic outflow is large. In fact,
it would overproduce the optical emission if all the energy is released in the optical
band. In the models discussed here, only a small fraction of the energy, $\sim10^{-2}$,
is released as synchrotron radiation, since electrons accelerated in residual collisions
cool mainly by IC scattering of the prompt GRB $\gamma$-rays. This has some important
implications to observations at high energy, $> 100$~MeV. Such observations are expected
to be useful in determining the bulk Lorentz factor of GRB outflows and the size of the
emitting region by detecting the high energy cutoff due to $\gamma\gamma$ absorption
\citep[e.g.,][]{Baring00,Lithwick01,LZ03}. The identification of this cutoff may be
complicated in the the presence of strong high energy emission from residual collisions,
that take place at large radii where the $\gamma\gamma$ optical depth is reduced.

A comment is in place here regarding some recent constraints on the size of the GRB
emission region, which were inferred from the early X-ray steep decay. Assuming that the
steep decay arises from emission by plasma lying away from our line of sight, large radii
were inferred, $R_{\rm em}>2t_{\rm decay}c/\theta_j^2\ga6\times10^{13}$cm for $t_{\rm
decay}\ga10^2$s and $\theta_j\la0.3$ \citep{Lazz06,Lyutikov06,Kumar07}. Such radii are
larger than typically predicted in internal shock models. It should be realized, however,
that if the off-the-line-of-sight emission explanation is adopted, the emission during
this phase should peak below the X-ray band, and should therefore arise in a region
different than that where $\gamma$-rays are produced. This is due to the fact that the
flat X-ray spectrum ($F_\nu\propto \nu^{0}$) seen in GRB spectra would imply a light
curve decay $\propto t^{-2}$ \citep[e.g.,][]{Kumar00}, much shallower than observed
\citep[e.g.,][]{Tagli05}. In fact, the spectra during the steep decay are soft,
suggesting that indeed the energy peak is below the X-ray band. Thus, if the steep decay
is due to off-the-line-of-sight emission, it should originate from a region lying at a
larger radius than that where $\gamma$-rays are produced, producing emission that peaks
below the X-ray band. Such emission may be produced, e.g, by residual collisions.

\acknowledgments

This research was supported in part by ISF and Minerva grants.

\end{document}